\documentclass{amsart} 
\usepackage{amsaddr}
\usepackage{graphicx}

\newcommand{\ket}[1]{\left| \, #1 \right\rangle}

\begin{document}
\title{The Early Days of Quantum Computation} 
\author{Peter W. Shor} 
\address{Department of Mathematics \\ Massachusetts Institute of Technology} 
\begin{abstract}
I recount some of my memories of the early development of quantum computation, including the discovery of the factoring algorithm,
of error correcting codes, and of fault tolerance.
\end{abstract}
\maketitle

This lecture was originally prepared for the occasion of the 40th anniversary of the Physics and Computation conference at Endicott House in 1981, so I thought I should start in 1981. I was a senior at Caltech then, and I must have been there when Feynman was preparing his keynote address for the Endicott House conference \cite{Feyn82}, which was one of the first times anybody thought carefully about quantum computation.  I didn’t hear anything about it when I was at Caltech, and in fact, I didn’t see Feynman’s paper until much later. But I do want to mention another lecture of his that I heard him give at Caltech, which shows that he was thinking about problems in the foundations of physics at the time.  

Feynman's talk was on negative probability. At the beginning of the talk, he explained that he had been looking at Bell’s theorem, which shows that quantum physics cannot be a local realistic hidden-variable theory.  What that means is that any interpretation of quantum mechanics either requires non-locality, or requires non-realism (here locality means that information can't travel faster than light, and realism means that things you can measure correspond to concrete properties of particles).  Feynman explained that what he had done was look carefully at the hypotheses that went into proving Bell’s theorem, to see whether there were any hidden assumptions. And indeed he found one -- the assumption was that all probabilities were between $0$ and~$1$. He reasoned that maybe there was a way around the EPR paradox if probabilities could be less than $0$ or larger than $1$, but that when you calculated any probability that you could actually observe, the calculation would add up these unrealistic probabilities and obtain a result between $0$ and~$1$. This isn't as outlandish as it may sound at first -- the Wigner function for the harmonic oscillator behaves in exactly this way, and Feynman remarked on that. He went on to show some things that he had figured out about negative probabilities; I don’t remember this part of the lecture very well.

Earlier, at a series of lectures in 1964 \cite{Feyn65}, Feynman had said 
\begin{quote}
I am going to tell you what nature behaves like. If you will simply admit that maybe she does behave like this, you will find her a delightful, entrancing thing. Do not keep saying to yourself, if you can possibly avoid it, `But how can it be like that?' because you will get `down the drain', into a blind alley from which nobody has escaped. Nobody knows how it can be like that.''
\end{quote}
But here, 15 years later, he’s investigating one possible reason for how it might be like that. So you see from this that Feynman didn’t take his own advice. Of course, maybe the advice wasn’t for tenured professors but for graduate students; for them, it’s probably very good advice -- finding new, significant results in the foundations of quantum mechanics is really hard, so it is not a good topic to give a graduate student to study. 

Feynman wrote down some of the ideas from the lecture a few years later, in 1983, in a paper called “Negative Probability” \cite{Feyn87} which wasn’t published until several years after that. The interesting thing about the paper is that it doesn’t mention Bell’s Theorem for the motivation. His paper first discusses Wigner functions, which do give negative values for probability. Feynman then generalizes Wigner functions to qubits in spin systems. And Feynman’s original motivation that he talked about in his lecture has been replaced by a motivation involving getting rid of the infinities in quantum field theory.  So presumably Feynman’s original idea didn’t work -- he wasn’t able to solve the EPR paradox using negative probabilities. I don’t know exactly what to make of the change in motivation. Did he think that trying to find a better way to understand Bell's Theorem was too disreputable a motivation because it involved the foundations of quantum mechanics, so he came up with a more respectable one?  Or was there some other reason?

There’s another interesting paper from this era that I want to mention. In 1985, David Deutsch wrote a paper on quantum Turing machines, quantum computing, and the quantum Church-Turing thesis. Deutsch’s motivation arises in part from the foundations of quantum mechanics \cite{Deut85} -- he asserts that with a quantum computer, it would be possible to test the Everett (many-worlds) interpretation of quantum mechanics. He says in the paper 
\begin{quote}
The intuitive explanation of these properties places an intolerable strain on all interpretations of quantum theory other than Everett’s.
\end{quote}
 Let me note that I disagree with him about this, although he hasn’t changed his opinion.  
  
 In 1985, Feynman wrote a followup \cite{Feyn86} to his 1982 paper, where he gave a much more detailed proposal of how one might build a quantum computer. This proposal still wasn't practical, though, because it requires exceedingly high precision in constructing the initial state and the transition amplitudes.

As an interesting side note, let me comment that both David Deutsch and Richard Feynman were thinking about quantum foundations when they were looking at quantum computing. My gut feeling is that this is significant. If you subscribe to David Mermin’s version of the Copenhagen interpretation --``shut up and calculate''-- then you avoid thinking about quantum weirdness, so maybe you also avoid thinking about possible uses for quantum weirdness.

At Caltech, I took a number of courses in physics, but majored in math. I went on to grad school in applied math at MIT, studying math that overlapped with computer science, and eventually got a job at Bell Labs doing math and computer science. The first time I heard of quantum information was when Charlie Bennett gave a talk at Bell Labs about the BB84 quantum key distribution protocol, invited by Bennett and Gilles Brassard \cite{BB84}. I don’t remember what year it was, but it must have been in the late 1980s.  I do remember that I was very intrigued by the talk, and thought for a little while about an open problem Charlie stated. The open problem was whether you could prove rigorously that BB84 was secure. Thinking about this, it wasn’t at all clear to me how you could formalize quantum mechanics into mathematics in a such a way as to give a rigorous proof that it was secure. So I gave up on this problem.

This is kind of ironic, because in 2000, after I knew much more about quantum computation, quantum information, and quantum error correcting codes, John Preskill and I were able to come up with a simple proof that BB84 was indeed secure \cite{PS00}. This was the third proof that BB84 was secure, but the first two proofs, by Meyers \cite{Mayers01} and by Biham et al. \cite{BBBvdGM02} were quite complicated. I read Mayers' proof carefully and realized that CSS codes were implicit in Mayers' proof, and that by using these codes, you could come up with a much simpler proof.

After Bennett's talk at Bell Labs, I didn’t think about quantum computing again until 1992, when Umesh Vazirani gave a talk at Bell Labs about his paper with Ethan Bernstein on quantum Turing machines. This paper was later published in STOC,  one of the two most prestigious theoretical computer science conferences, in June 1993 \cite{BV93}. I was really intrigued by that talk, and I probably understood it better than other computer scientists because of the amount of physics I’d taken in college. Bernstein and Vazirani gave a rather contrived problem which quantum computers seemed to do better at, but I wasn’t really satisfied with this problem. So I started thinking about whether you could solve a real problem more efficiently with a quantum computer. 

I didn’t really get anywhere until I saw Dan Simon’s paper \cite{Simons94} where he solves a contrived problem, “Simon’s problem”, using a quantum computer algorithm that works much better than the best classical algorithm for this problem. I saw this paper because I was on the program committee for the 1994 STOC conference, where we rejected the paper (despite my arguing for it). I’ve been told that Dan Simon started out by trying to prove that digital computers weren’t any more powerful than quantum computers, but he eventually found a problem that showed the opposite.

Simon had used the Fourier transform over a binary vector space to find the period of a function over this vector space. I knew that Fourier transforms were useful for finding periods, and I knew that the discrete log problem was related to periodicity, so I started looking for ways to solve the discrete log problem on a quantum computer.  The discrete logarithm problem was a famous problem that, if you could solve it, you could break some public key cryptosystems. 

\begin{figure}[t]
\includegraphics[width=2.5in]{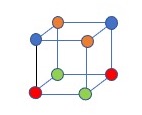}\quad\raisebox{.14in}{\includegraphics[width=2in]{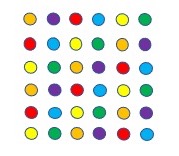} } \\
\hspace*{.75in} {\LARGE Hypercube} \hspace*{1.45in} {\LARGE Torus} \\
\caption{Left: An example of period-finding on the vertices of a hypercube. Here, the period is $(1,1,0)$. Right: An example
of period-finding on a 2-dimensional torus. Here, the period is $(2,1)$. \\  \label{cube-torus}}
\end{figure}

The difference between period-finding for Simon's algorithm and for the discrete logarithm problem is that for Simon's problem, you need to find the period of a function on an $n$-dimensional hypercube, while for the discrete logarithm problem, you need to find the period on a $P-1 \times P-1$ torus. (See Fig.~\ref{cube-torus}.) It wasn't at all obvious how to get from period-finding on a high-dimensional hypercube to period-finding on a large torus.
You need different quantum Fourier transforms for these. The first thing I was able to do was to solve the discrete log problem in a special case that was already solvable on a classical computer in polynomial time. While this didn’t really provide anything new, the algorithm was completely different from the classical algorithm and it gave me enough encouragement that I kept on working on it, and eventually solved the general case. The insight I needed for going from the special case to the general case was realizing that for the general case, I didn't need to take the Fourier transform modulo $P-1$; it was sufficient to take it modulo a number that was somewhat larger than $P-1$. 

I didn’t tell anybody I was working on the discrete log problem, because I thought it was such a long shot. When I solved it, in April 1994, I told a few people I knew and worked with, including my boss David Johnson, and a colleague at Bell Labs, Jeff Lagarias, who verified that my algorithm was correct. Actually, it wasn’t quite correct -- Jeff found a small error that was easily fixable. After that, David Johnson suggested I give a talk on it in Henry Landau’s seminar.

Henry Landau’s seminar was held every Tuesday, and there was a very active audience -- they constantly interrupted the speaker with questions, so it was a little intimidating. I remember that on one occasion, the speaker never got past his second slide. Anyway, I gave a talk on how to solve discrete logarithms on a quantum computer, and it went well. Later that week, I was able to solve the factoring problem as well. There’s a strange relation between discrete log and factoring. There’s no formula for taking an algorithm for one of these problems and applying it to the other. However, any time somebody has found an improved algorithm for one of them, people have reasonably quickly come up with a similar solution for the other one. This case was no different -- knowing the discrete log algorithm, it was fairly easy to find a factoring algorithm.

That weekend, when I was at home with a bad cold, Umesh Vazirani called me up and said “I hear that you can factor efficiently with a quantum computer.” This was surprising for a couple of reasons. First, it showed that rumors of my Tuesday talk had spread until they reached Umesh. Second, the talk had been about the discrete log algorithm, but by the time the rumors reached Umesh, they had changed into factoring (this is a little bit like the game of Telephone, also called Whisper Down the Lane, where a whispered message gets passed from one player to the next and changes along the way). But luckily, I had solved the factoring problem in the meantime, so rather than having to tell Umesh that his information was wrong, I was able to explain the factoring algorithm to him.

After that, the news spread like wildfire. There was a Columbia University Theory Day (on theoretical computer science) later in April. Charlie Bennett, John Smolin, and I arranged to meet there, and I explained the algorithm to them. If I remember correctly, Bennett explained his recent teleportation protocol \cite{BBCJPW93} and the recent Elitzur-Vaidman bomb-testing protocol \cite{EV93} to me. Quantum information was clearly a subject whose time had come.

I was invited to give a last-minute special talk at the Algorithmic Number Theory Symposium at Cornell at early May. There was a conference at the Santa Fe Institute on quantum mechanics in mid-May; I couldn’t go, but Umesh presented my algorithm at this conference.  Artur Ekert and David Jozsa wrote up a paper explaining my algorithm to physicists \cite{EJ96} and Artur Ekert gave a talk about it in the Atomic Physics conference in Boulder, Colorado that August. This is where a number of physicists heard about it, as you heard in David Wineland's talk.

During this period, I got requests for interviews from magazines, and many requests for a copy of my paper. (I started sending it out before it was completely finished, which may have been a mistake, as I was answering questions about mistakes in early drafts long after I’d fixed them in the paper.) There was a conference arranged by NIST in Gaithersurg, MD, that August, which was arranged because of the discovery of the factoring algorithm, and whose focus was quantum computation. There was a workshop on Quantum Computation in Turin, Italy, in October, which was actually the second or third in the series. This conference started with a handful of people, and grew every year until it outgrew the conference facilities at the Villa Gualino, at which point the conference was discontinued. Physcomp ’94 in Texas was a follow-up to the 1981 Endicott House conference (there were three of these follow-ups, in ’92, ’94, and ’96) and I gave a talk there. Then I presented my paper at FOCS in Santa Fe, NM just a few days later, where it was published in the proceedings \cite{Shor94}. 

There was one big objection to quantum computation, which Rolf Landauer had already raised at the conference at the Santa Fe Institute in May. It didn’t look like quantum computers could provide fault tolerance. And without fault tolerance, if you need to do $N$ steps on a quantum computer, you need to make each step accurate to $1/N$. For $N$ large enough, like a billion (which is something like what you need to factor cryptographically significant large numbers) this was seen as absolutely impossible by experimental physicists. 

There were two main quantum mechanical principles seen as obstacles for error correction, the Heisenberg Uncertainty Principle and the Quantum No-Cloning theorem. The Heisenberg uncertain principle says that you cannot measure the state of a quantum computer completely. The quantum no-cloning theorem says that you cannot make a copy of an unknown quantum state. 

Suppose you’re building a classical computer out of unreliable components, and you want to make it fault-tolerant. There are a bunch of techniques you can use. One is checkpointing -- you write down the state of your computation periodically, and if the computation derails at some point, you don’t have to start over at the beginning, you can just start over at the checkpoint. Another technique is error-correcting codes. These use redundancy to help you fix any bits that get corrupted while in memory. Finally, the last technique is massive redundancy. You keep multiple copies of every bit in your computation, and continually compare them against each other and fix the ones that are wrong. Massive redundancy appeared to be the most powerful of these techniques, and it was investigated by von Neumann in 1956 \cite{vN56}.

The problem is that the quantum no-cloning theorems seem to say that all of these are impossible. For the checkpointing, you can’t write down the state of your computation and keep on computing -- that’s making a copy. For massive redundancy, fixing an error involves making a copy -- if you have four good copies of your computation and one bad copy, then getting five good copies from this is something the no-cloning theorem says is impossible.

Fortunately, even though it looks like they require redundancy, error correcting codes still work. Although I hadn't studied them in school,  I knew something about error-correcting codes from being in the Math Center at Bell Labs. The simplest classical error correcting code is the repetition code, where you just make several copies of your bit, and you use majority vote to fix any errors. The shortest code this works for is the three-bit code (since you need a majority). You can do this for a quantum code as well. 
This code encodes one qubit into three qubits as follows:
\begin{align*}
\ket{0}_L& = \ket{000} , \\
\ket{1}_L &= \ket{111}.
\end{align*}
Here, $\ket{}_L$ stands for the logical qubit, which is encoded into three physical qubits. This code corrects bit errors, but it makes phase errors three times as likely.

What I realized is that there’s a transform on quantum codes, which we now call the Hadamard transformation, that takes bit errors to phase errors and vice versa. 
This transformation is $H = \frac{1}{\sqrt{2}} \left( \begin{array}{rr} 1 & 1 \\ 1 & -1 \end{array}\right) $.
If you apply this transformation, you get a phase-error correcting code, which corrects phase errors, but makes bit errors three times as likely. 
this code is 
\begin{align*}
\ket{0}_L &= \frac{1}{\sqrt{8}} \left( \ket{0} + \ket{1} \right)^{\otimes 3} , \\
\ket{1}_L &= \frac{1}{\sqrt{8}} \left( \ket{0} - \ket{1} \right)^{\otimes 3} .
\end{align*}

You can combine these two codes by a process called concatenation, which is a very important technique from classical coding theory: First, you encode the qubit you want to protect
using one of these quantum codes. You then encode each qubit of the resulting state with the other code. When you combine them this way, 
you get the following 9-qubit code that corrects both bit errors and phase errors: 
\begin{align*}
\ket{0}_L &= \frac{1}{\sqrt{8}} \left( \ket{000} + \ket{111} \right)^{\otimes 3} , \\
\ket{1}_L &= \frac{1}{\sqrt{8}} \left( \ket{000} - \ket{111} \right)^{\otimes 3} .
\end{align*}
So this is how I discovered the 9-qubit code.

Sometime after I wrote the paper, I discovered that the three-qubit bit-error correcting code had already been discovered by Asher Peres \cite{Peres85}. It appeared in a paper that Asher wrote in 1985 on quantum computers (possibly inspired by work that came out of the 1981 Endicott House conference). Asher wasn’t trying to do quantum computation, just classical computation using quantum components, so he didn’t need to worry about phase errors. But he did give an excellent analysis of the three-qubit bit error correcting code. 

Classically, the repetition code had been known, at least informally, for millennia. However, more complicated classical error correction codes, which are much more efficient that the repetition code, had only been discovered less than fifty years earlier; the first of these codes were discovered by Richard Hamming \cite{Hamm50}. So proceeding by analogy, I decided to look for more complicated quantum error correcting codes. I started playing around with the classical seven-bit Hamming code -- the simplest classical code more complicated than repetition codes -- and discovered the quantum version of it, which encodes one qubit into seven qubits and corrects one error. The key to this again was the Hadamard transformation, which takes bit errors to phase errors and vice versa. The classical Hamming code corrects bit errors. However, if you make a superposition of its codewords in the right way, it is invariant under the Hadamard transformation, and so can correct both bit errors and phase errors. This gives the seven-qubit quantum Hamming code:
\begin{align*}
\ket{0}_L =  \frac{1}{\sqrt{8}} \big( &\ket{0000000} + \ket{1111000} + \ket{1100110} + \ket{0011110} \\
+&\ket{1010101} + \ket{0101101} + \ket{0110011} + \ket{1001011}\big) , \\
\ket{1}_L =  \frac{1}{\sqrt{8}} \big( &\ket{1111111} + \ket{0000111} + \ket{0011001} + \ket{1100001} \\
+&\ket{0101010} + \ket{1010010} + \ket{1001100} + \ket{0110100}\big) .
\end{align*}
I showed this construction to Rob Calderbank, and we figured out a generalization to a large class of quantum error correcting codes that are constructed by combining two classical codes that are weakly dual to each other \cite{CS96}. Andrew Steane discovered the quantum Hamming code and this construction at around the same time \cite{Steane96a,Steane96b}, so these are now called CSS codes after their discoverers.

After these developments, people started looking for other quantum error correcting codes. Two groups put the question on a computer, one at Los Alamos National Labs and the other at IBM, and found a five-qubit code. The two five-qubit codes looked quite different, but you could apply a number of transformations and show that they were really the same. And while it was apparent from looking at them that they had quite a bit of structure, it wasn’t clear what exactly this structure was.

When I was trying to figure out the structure of the code, the first thing I decided to do was find its symmetry group. I asked Neil Sloane how you would find the symmetry group, and he pointed me to some software -- MAGMA, to be exact -- and gave me a sample MAGMA program that he had written to find out the size of a group he was working on for his research. The software showed that my group was the same size, 5160960, as his group. Not only that, but looking more closely, it turned out to be the same group, and further that there was a deep connection between the two problems. This led us to discover the theory of stabilizer codes \cite{CRSS97,CRSS98} (which Daniel Gottesman \cite{Gott96} simultaneously discovered). 

There were a couple of other interesting developments during this period. Alexei Kitaev heard about the factoring result, but being in Russia, he somehow couldn’t get ahold of the paper. So he figured out a different proof of the result, which gave us the phase estimation algorithm \cite{Kitaev95}. And Lov Grover, at Bell Labs, discovered a quantum search algorithm that does quadratically better than the best classical search algorithm \cite{Grover97}.

The last thing I want to talk about is fault tolerance. To be able to build a quantum computer, it’s not enough to be able to correct errors with noiseless gates; you need to be able to correct errors using noisy gates. This means you have to correct the errors faster than you introduce new ones. Von Neumann showed how to do this with classical noisy gates in 1956 \cite{vN56}.  But it’s trickier with qubits -- you need to figure out how to perform operations on encoded qubits without ever decoding them, because as soon as you decode your logical qubits, you potentially expose them to error. I realized that this was fairly straightforward for gates in the Clifford group, because for a certain class of CSS codes
you could perform these gates transversally, i.e. with circuits where the $i$th qubit encoding a logical qubit only interacts with the $i$th qubits of other logical qubits. This separates the $i$th qubits of codewords from the $j$th qubits, so the errors don't spread very far. However, this only works for gates in the Clifford group, and Clifford group gates do not allow you to do universal computation. In fact, if your quantum circuit has only gates in the Clifford group, it can be simulated by a classical computer.

How can we perform non-Clifford gates on encoded qubits? It turns out that we  only need to figure out how to implement a single gate that is not in the Clifford group. 
The first thing I tried was to implement the gate 
\[
R_{\frac{2 \pi}{3}} = \left(\begin{array}{cc} -\frac{1}{2} &-\frac{\sqrt{3}}{2} \\ 
\phantom{-}\frac{\sqrt{3}}{2} &  -\frac{1}{2} \end{array} \right)\]  on encoded qubits.
What happens when we apply this gate to an arbitrary encoded qubit, $\ket{\psi}_L = \alpha \ket{0}_L + \beta \ket{1}_L$? We get the state
\[
R_{\frac{2 \pi}{3}}   \ket{\psi}_L = - \frac{1}{2} \alpha \ket{0}_L + \frac{\sqrt{3}}{2} \alpha\ket{0}_L -\frac{\sqrt{3}}{2} \beta\ket{1}_L - \frac{1}{2} \beta \ket{1}_L.
\]
Compare this with the state
\[
\bigg(  -\frac{1}{2} \ket{0}_L + \frac{\sqrt{3}}{2} \ket{1}_L \bigg) \ket{\psi}_L =
 - \frac{1}{2} \alpha \ket{00}_L + \frac{\sqrt{3}}{2}\alpha\ket{10}_L  - \frac{1}{2} \beta\ket{01}_L  +\frac{\sqrt{3 }}{2} \beta\ket{11}_L.
 \]
How can we change this state to the desired state, $R_{\frac{2 \pi}{3}}   \ket{\psi}_L $? We can apply a controlled $\sigma_z$ gate to change the sign on the state $\ket{11}_L$, and can then
apply a CNOT, or controlled $\sigma_x$ gate from the second qubit to the first qubit. (Controlled Pauli gates are in the Clifford group, so we can apply them fault-tolerantly.) This gives us the state
\[
 - \frac{1}{2} \alpha \ket{00}_L + \frac{\sqrt{3}}{2}\alpha\ket{10}_L  -\frac{\sqrt{3}}{2} \beta\ket{01}_L- \frac{1}{2} \beta\ket{11}_L  .
\]
Now, if we measure the first encoded qubit in the basis $\frac{1}{\sqrt{2}}(\ket{0} \pm \ket{1})$ and we get the outcome
 $\ket{+} = \frac{1}{\sqrt{2}}(\ket{0}  + \ket{1})$, we have the desired state $R_{\frac{2 \pi}{3}}   \ket{\psi}_L $.
If, on the other hand, we get the outcome $\ket{-}$, we have the state $R_{-\frac{2 \pi}{3}}   \ket{\psi}_L $. We thus have found a procedure where we can rotate the state
by $\frac{2\pi}{3}$ either clockwise or counterclockwise, with a probabiity of $\frac{1}{2}$ of going either direction. However, since two clockwise rotations by this angle make a counterclockwise
rotation, we can repeat this procedure until we have the state $R_{\frac{2 \pi}{3}}   \ket{\psi} $, and the expected number of rotations we need to make is only $2$.

The difficulty with the above idea is that it wasn't clear how to prepare an encoded state $-\frac{1}{2} \ket{0}_L + \frac{\sqrt{3}}{2} \ket{1}_L $. After some time and effort, I figured out how to use a similar
construction to build a Toffoli gate. Instead of using the ancillary state $-\frac{1}{2} \ket{0}_L + \frac{\sqrt{3}}{2} \ket{1}_L$, we use the state $\frac{1}{2}(\ket{000}_L + \ket{100}_L + \ket{010}_L+ \ket{111}_L)$.
The key to preparing this ancillary state was to use the cat state $\frac{1}{\sqrt{2}} (\ket{000\ldots0} + \ket{111\ldots1})$. You can check this state to make sure that it is a superposition of states containing mostly $0$'s and states containing mostly $1$'s. You cannot easily check the phase on the superposition; however, if you build your circuit carefully, an error in this phase only results in correctible errors in the the quantum code, which can be handled by the error correction circuits.

My paper did not give the result for fault-tolerance that I wanted to prove, which was a {\em threshold theorem}. The threshold theorem says that if you have a sufficiently low constant error rate, you can take any quantum circuit and build a fault-tolerant version of it that takes no more than polynomial extra overhead. One deficiency of my paper was that it only showed how to perform a limited set of gates. I showed how to perform all Clifford gates and the Toffoli gate. There are a few more gates that you can find exact constructions for. However, by the nature of quantum fault tolerance, for any fault-tolerant protocol, there are only a discrete set of gates that it can perform. This is because if it could fault-tolerantly implement a family of gates that depend on a continuous parameter, you couldn't discriminate between two close value of this continuous parameter. Thus, what you need to do is find a discrete set of gates that will give a good approximation of any unitary transform on some small number of qubits. The Solovay-Kitaev theorem \cite{Kitaev97} shows that this is possible. In fact, the theorem shows that if you have any finite set of quantum gates in SU($k$) which generate a group dense in SU($k$), then any gate in SU($k$) can be approximated well by a relatively short sequence of gates from this set. Using this, you can show that if you have fault-tolerant operations for any set of gates that generate a dense group in SU(k), you can use these approximations to build a fault-tolerant circuit that approximates any circuit sufficiently well, with only polylogarithmic overhead.

My paper also didn't show that quantum computers could be made completely fault-tolerant. It showed that if you had a quantum hardware with an error rate of $\epsilon$, you could perform on the order 
$e^{c \sqrt[4]{1/\epsilon}}$ gates with an overall small probability of error, for some constant $c$. However, what I really wanted to show is that there is some threshold, so if the error $\epsilon$ is below this threshold, fault tolerance will work
for arbitrarily long computations.
Two groups of researchers \cite{AB-O97, AB-O08, KLZ98} were able to show this result by concatenating my construction with itself many times. To compute for $n$ steps, you need $\log \log n$ levels of concatenation, and for this you pay a polylogarithmic overhead. Alexei Kitaev \cite{Kitaev03} found a different way to perform fault-tolerant quantum computation, by using topological codes.
The discovery of the threshold theorem showed that quantum computers might be technologically possible (although still very difficult), and led to an outburst of research into ways that they might actually be built.

\bibliographystyle{amsplain} 

\providecommand{\bysame}{\leavevmode\hbox to3em{\hrulefill}\thinspace}
\providecommand{\MR}{\relax\ifhmode\unskip\space\fi MR }
\providecommand{\MRhref}[2]{%
  \href{http://www.ams.org/mathscinet-getitem?mr=#1}{#2}
}
\providecommand{\href}[2]{#2}

\end{document}